\documentclass[12pt,letterpaper]{amsart}
\usepackage{amssymb,latexsym, amsmath, amsxtra}
\usepackage[dvips]{graphics}

\theoremstyle{plain}
\newtheorem{theorem}{Theorem}[section]

\newtheorem{lemma}[theorem]{Lemma}

\theoremstyle{definition}

\theoremstyle{remark}

\numberwithin{equation}{section}

\newcommand{\R}{\mathbb R}

\newcommand{\C}{\mathbb C}

\renewcommand{\Re}{\mathrm{Re}}
\renewcommand{\Im}{\mathrm{Im}}

\def\({\left(}
\def\){\right)}

\newcommand{\ontop}[2]{\genfrac{}{}{0pt}{}{#1}{#2}}

\begin{document}
\title[Random polynomials and $L$-functions]
{Random polynomials, random matrices and $L$-functions}
\author{David W Farmer, Francesco Mezzadri, and Nina C Snaith }

\thanks{Research supported by the
American Institute of Mathematics and the Focused Research Group
grant (0244660) from the NSF. This work was started during the
program on Random Matrix Applications in Number Theory at the
Isaac Newton Institute for Mathematical Sciences. The second and
third authors were also supported by a Royal Society Dorothy
Hodgkin Fellowship, and the third author was partially supported
by EPSRC}

\date{\today}
\thispagestyle{empty}
\vspace{.5cm}
\begin{abstract}
We show that the Circular Orthogonal Ensemble of random matrices
arises naturally from a family of random polynomials.  This
sheds light on the appearance of random matrix statistics in the
zeros of the Riemann zeta-function.
\end{abstract}

\address{
{\parskip 0pt American Institute of Mathematics\endgraf 360
Portage Ave.\endgraf Palo Alto, CA 94306\endgraf USA \endgraf
farmer@aimath.org\endgraf \null \null School of
Mathematics\endgraf University of Bristol \endgraf Bristol BS8
1TW\endgraf United Kingdom\endgraf
f.mezzadri@bristol.ac.uk\endgraf n.c.snaith@bristol.ac.uk\endgraf
}
  }

\maketitle

\section{Introduction}

The statistics of eigenvalues of unitary matrices, chosen
uniformly with respect to Haar measure on~$U(N)$, are observed to
closely match the statistics of zeros of the Riemann zeta-function
and other
$L$-functions~\cite{kn:mont73,kn:odlyzko89,kn:hejhal94,kn:rudsar}.
In addition, eigenvalues of other
compact classical matrix groups
give a good
model of the zeros of various families of
$L$-functions~\cite{kn:katzsarnak99b,kn:katzsarnak99a,kn:rub98,kn:ILS99,kn:farlem05}.
Furthermore, the characteristic polynomials of the matrices
provide a good model of the $L$-functions
themselves~\cite{kn:keasna00a,kn:confar00,kn:keasna00b,kn:cfkrs,kn:hughes00}.

The $L$-functions studied in number theory are Dirichlet series
having a functional equation and an Euler product.  In this paper
we are concerned with a wider class of Dirichlet series which have
a functional equation but do not have an Euler product (see, for
example,~\cite{kn:fk,kn:hejhal87}). It has been suggested that
such functions can be modelled by random self-reciprocal
polynomials~\cite{kn:fk}. In Appendix A we discuss the example of
Epstein zeta functions. It is also possible to create Dirichlet
series with a functional equation but no Euler product from Maass
forms on a nonarithmetic group~\cite{kn:farlem05b}. Since these
functions do not have an Euler product, they are not expected to
satisfy the Riemann hypothesis. However, it is possible that
occasionally such functions will satisfy the Riemann hypotheses
or, as is more likely, will have a large number of consecutive
zeros, say the first 100 trillion of them, on the critical line.

This paper is
motivated by the following questions:  do the Riemann
zeta-function and the other $L$-functions of number theory
behave differently than random Dirichlet series with functional
equation which just happen to satisfy the Riemann hypothesis?
That is, does the Euler product have any effect on
the zeros beyond forcing them to be on the critical line?
Our results suggest that the answer is `yes', and the Euler
product also has an effect on the local statistics of the zeros.

\subsection{Random polynomials}

In random matrix theory, the characteristic polynomial of a random
matrix can be viewed as a random polynomial where the randomness
is explicitly encoded in the zeros. For example, consider the Weyl
integration formula for the classical compact
groups~\cite{kn:weyl,kn:katzsarnak99a} or the $\beta=1$, $2$, $4$
ensembles of random matrix theory~\cite{kn:mehta}. On the other
hand, usually in the study of random polynomials (for a review of
the subject see, e.g., Farahmand~\cite{kn:farahmand}) the
randomness is explicitly encoded in the coefficients of the
polynomials.  We will show that there is a simple but surprising
connection between these two perspectives, and we believe this
connection is relevant to the appearance of random matrix
statistics in the zeros of the Riemann zeta-function and other
$L$-functions.

The polynomials we consider are of the form
\begin{equation}\label{eqn:defofpoly}
f(z)=z^N + a_1 z^{N-1} + \cdots + a_N
\end{equation}
with $|a_N|=1$ which have the symmetry
\begin{equation}
\label{eqn:selfrecip} f(z)= a_N z^N \overline{f}\(\frac{1}{z}\),
\end{equation}
where $\overline f(z):=\overline{f(\overline z)}$. Such
polynomials are called \emph{self-reciprocal}.
Equation~\eqref{eqn:selfrecip} ensures that the zeros of $f$ occur
either on the unit circle or in pairs located symmetrically with
respect to the unit circle. The symmetry~\eqref{eqn:selfrecip} is
analogous to the functional equation of $L$-functions which arise
in number theory, see \eqref{eqn:asymfe} below.  In terms of the
coefficients $a_j$ the functional equation~\eqref{eqn:selfrecip}
translates into the relation
\begin{equation}
a_{N-j} = a_N \overline{a}_j.
\end{equation}

The change of variables $z=e^{i x}$ in the function
$a_N^{-\frac12}z^{-N/2} f(z)$ converts a self-reciprocal algebraic
polynomial into a \emph{real} trigonometric polynomial
\begin{equation}
\label{realtrig} \sum_{0\le n\le N/2} c_n
\cos\left(\left(N/2-n\right) x\right) + d_n \sin\left(\left(N/2 -
n\right) x\right) ,
\end{equation}
 where
$c_n$ and $d_n$ are real. And if $f$ has real coefficients then
the associated trigonometric polynomial is an even function
of~$x$, having only cosine terms in its expansion. The roots of
these trigonometric polynomials are either on the real line or in
complex conjugate pairs.  This is somewhat closer to the symmetry
of an $L$-function, but in this paper we phrase everything in
terms of algebraic polynomials~\eqref{eqn:defofpoly} to emphasize
the comparison with characteristic polynomials of random matrices.

For even trigonometric polynomials, whose coefficients are
independent standard normal random variables,
Dunnage~\cite{kn:dunn66} discovered that the expected number of
real zeros is given by
\begin{equation}
\label{nrealzeros}
\frac{2N}{\sqrt{3}} + O\left(N^{11/13}(\log
N)^{3/13}\right).
\end{equation}
Bogomolny~{\it et al.}~\cite{kn:bbl96} extended Dunnage's result
and studied the average fraction of roots of the
polynomial~\eqref{eqn:defofpoly} on the unit circle when $a_N = 1$
and the coefficients $a_j$ (for $1 \le j < N/2$) are complex
normal random variables with mean zero and whose standard
deviation $\sigma$ varies with $N$ as
\begin{equation}
\sigma = \epsilon/\sqrt{N}.
\end{equation}
They discovered that in the limit $\epsilon \rightarrow 0$,
\emph{i.e.} when all the coefficients have a narrow distribution
centered around zero, there is a small neighbourhood of $\epsilon$
where the average fraction of roots lying on the unit circle is
one. Instead, when the coefficients have a broad distribution,
\emph{i.e.} in the limit $\epsilon \rightarrow \infty$, the
average fraction of zeros on the unit circle tends to
$1/\sqrt{3}$.  They also computed the two-point correlation
function of the zeros on the unit circle, which for short
distances grows linearly.

\subsection{$L$-functions}

An $L$-function is a Dirichlet series,
\begin{equation}
L(s)=\sum_{n=1}^\infty \frac{a_n}{n^s},
\end{equation}
with $a_n = O_\epsilon(n^\epsilon)$ for every $\epsilon>0$, which
has an analytic continuation to the complex plane (except for
possible poles on the line $\Re s=1$) along with two additional
properties. First, it has a  functional equation
\begin{equation}
 \xi_L(s) :=  \gamma_L(s)L(s) = \varepsilon
\overline{\xi_L}(1-s),
\end{equation}
with $|\varepsilon|=1$ and $\gamma_L$ of the form
\begin{equation}
\gamma_L(s)= P(s) Q^s \prod_{j=1}^w \Gamma(w_j s+ \mu_j) ,
\label{eqn:gammafactors}
\end{equation}
where $Q>0$, $w_j >0$, $\Re \mu_j \ge 0$, and $P$ is a
polynomial whose only zeros in $\sigma>0$ are at the poles of
$L(s)$.
Second, it has an Euler product representation of the form
\begin{equation}
L(s)=\prod_p L_p(1/p^s),
\end{equation}
where the product is over the primes~$p$, and
\begin{equation}
L_p(1/p^s)=\sum_{k=0}^\infty \frac{a_{p^k}}{p^{k s}} =
\exp\left(\sum_{k=1}^\infty \frac{b_{p^k}}{p^{k s}} \right) ,
\end{equation}
where $b_n =O(n^\theta)$ with $\theta <\frac12$. It is conjectured
that such $L$-functions satisfy the \emph{Riemann hypothesis},
which is the assertion that the nontrivial zeros lie on the
critical line $\Re(s)=\frac12$.

The functional equation is a symmetry with respect to
the line $\Re(s)=\frac12$, so the nontrivial zeros of $L$
are on the line $\Re(s)=\frac12$, or they are located
symmetrically on either side of it.
To show the analogy with self-reciprocal polynomials~\eqref{eqn:selfrecip}
it is more convenient to write the functional equation in
asymmetric form:
\begin{equation}\label{eqn:asymfe}
L(s)=\varepsilon X_L(s) \overline{L}(1-s),
\end{equation}
where $ \displaystyle X_L(s) =
\overline{\gamma_L}(1-s)/\gamma_L(s)$. Note that $| X_L|=1$ on the
$\frac12$-line, that is, on the line of symmetry of the
$L$-function.  Thus, there is a perfect analogy between the
self-reciprocal property of $f$ and the functional equation
of~$L$.

\subsection{Results and discussion}

For Dirichlet series with a functional equation but no Euler
product, it is not expected that all zeros lie on the critical
line. However, it is possible that even if such a Dirichlet series
does not satisfy the Riemann hypothesis, it could have, say, its
first 100 trillion zeros on the critical line.  Our motivating
question is this: is the Riemann zeta-function, with its Euler
product, distinguishable from a random Dirichlet series with
functional equation that just happens to have its first 100
trillion zeros on the critical line? By considering the analogous
case of random self-reciprocal polynomials, we suggest that the
answer is `yes'.

Let us consider the space $\mathcal{R}_N$ of all the
self-reciprocal polynomials of degree $N$. Given a nonvanishing
probability distribution on $\mathcal{R}_N$ (meaning that nonempty open
sets have positive measure) the subset
$\mathcal{C}_N$ of those polynomials which happen to have all of
their zeros on the unit circle has positive measure. The proof is
quite simple. The zeros of a self-reciprocal polynomial lie either
on the unit circle or in pairs symmetric with respect to the unit
circle, and they are continuous functions of the coefficients. Two
zeros on the circle must ``collide'' in order to move off the
circle, so there is a small open neighborhood of coefficients in
which the zeros remain on the circle.  This neighbourhood has
positive measure. Therefore, the restriction to $\mathcal{C}_N$ of
the measure on $\mathcal{R}_N$ is unique and well defined.  Such a
restriction is then made into a probability measure by
renormalizing the volume of $\mathcal{C}_N$ to one.

Our interest is mainly concentrated on the distribution of the
roots of those polynomials whose zeros lie all on the unit circle.
Therefore,  our approach will be to define a joint probability
density function on the coefficients of the polynomials in
$\mathcal{R}_N$ and to study its restriction to $\mathcal{C}_N$.
The coefficients $a_n$ of polynomials with all their zeros on the
circle are bounded in magnitude by $\binom{N}{n}$, therefore not
only has $\mathcal{C}_N$  positive measure, but it is also
compact. Thus, the most natural choice is to put a distribution on
the coefficients $a_n$ which is uniform on a bounded disk
containing $\mathcal{C}_N$ and zero outside. The following theorem
gives the joint probability density function for the roots of such
polynomials. Petersen and Sinclair~\cite{PS} have found some
interesting geometric properties of the coefficients of these
polynomials.  They have also proved independently a result very
similar to Theorem \ref{thm:complexcase}.  In their work (Lemma
4.1 of \cite{PS}) they fix the coefficient $a_N$, rather than
allowing it to vary, but they arrive at the same Vandermonde form
for the Jacobian.

In what follows we denote by $\Delta(x_1,\ldots,x_N)$ the
Vandermonde determinant, i.e.
\begin{equation}\label{eq:vand}
\Delta(x_1,\ldots,x_N)= \prod_{j < k}(x_k - x_j),
\end{equation}
and we denote by $e_n$ the $n$th elementary symmetric function
\begin{equation}
e_n(x_1,\ldots,x_N)= \sum_{1\le i_1<\cdots<i_n\le N} x_{i_1}\cdots x_{i_n} .
\end{equation}

\begin{theorem}
\label{thm:complexcase}
Suppose $N$ is odd.
 Consider random monic polynomials
$z^N + \sum_{n=1}^{N} a_n z^{N-n}$ satisfying the self-reciprocal
property $a_{N-n}=a_N \overline{a}_n$, with
$a_1,\ldots,a_{(N-1)/2}$ chosen independently and uniformly in
$|a_n|\le \binom{N}{n}$, and with $\phi$ chosen uniformly in
$[0,2\pi)$, where $a_N=e^{i\phi}$, and restrict to those
polynomials having all zeros on the unit circle. The joint
probability density function of the set of zeros
$e^{i\delta_1},\ldots,e^{i\delta_N}$ is given, up to a
normalization constant, by
\begin{equation}
\label{oddcasef}
|\Delta(e^{i\delta_1},\ldots,e^{i\delta_N})|=\prod_{j<k}|e^{i\delta_k}-e^{i\delta_j}|.
\end{equation}

For $N$ even, consider self-reciprocal random monic polynomials
with $a_1,\ldots,a_{N/2}$ chosen independently and uniformly in
$|a_n|\le \binom{N}{n}$, and restrict to those polynomials having
all zeros on the unit circle.  The joint probability density
function for the set of zeros $e^{i\delta_1},\ldots,e^{i\delta_N}$
is given, up to a normalization constant, by
\begin{equation}
\label{evencasef}
|e_{N/2}(e^{i\delta_1},\ldots,e^{i\delta_N})\Delta(e^{i\delta_1},\ldots,e^{i\delta_N})|=
|e_{N/2}(e^{i\delta_1},\ldots,e^{i\delta_N})|\prod_{j<k}|e^{i\delta_k}-e^{i\delta_j}|,
\end{equation}
where $e_{N/2}(e^{i\delta_1},\ldots,e^{i\delta_N})$ is the $N/2$
elementary symmetric function in the variables
$e^{i\delta_1},\ldots,e^{i\delta_N}$ and is equal to $(-1)^{N/2}$
times the coefficient $a_{N/2}$.

In particular, the joint probability density function for odd $N$
is the same as that for eigenvalues of a randomly chosen matrix in
the Circular Orthogonal Ensemble $COE(N)$.
\end{theorem}

Some aspects of this result can be derived from~\cite{SHW} but
they approach the subject from a different perspective.

The theorem suggests that if a  Dirichlet series with functional
equation is chosen at random, and all of the zeros in a particular
interval of the critical strip happen to lie on the critical line,
then those zeros should have similar statistics to those of
eigenvalues of matrices from the COE.  The COE is the symmetric
space $U(N)/O(N)$  with the measure induced from Haar measure on
$U(N)$.  Thus the joint probability density function for the
eigenvalues is
\begin{equation}
\frac{1}{(4\sqrt{\pi})^N\Gamma(1 + N/2)}
|\Delta(e^{i\delta_1},\ldots,e^{i\delta_N})|=
\frac{1}{(4\sqrt{\pi})^N\Gamma(1 +
N/2)}\prod_{j<k}|e^{i\delta_k}-e^{i\delta_j}|.
\end{equation}
 All numerical calculations
of zeros of $L$-functions having an Euler product show the
statistics of the CUE (the group $U(N)$ with Haar measure), so
this suggests that the $L$-functions from number theory are not
typical Dirichlet series with a functional equation.

Thus, the Euler product which is considered a necessary condition
for a Dirichlet series with functional equation to satisfy
the Riemann hypothesis does more then just force the zeros onto
the critical line.  The Euler product fundamentally changes the
nature of the spacing of the zeros, in particular changing the
linear repulsion of zeros of random polynomials and the COE
into the quadratic repulsion of the CUE, whose joint
probability density function is given by
\begin{equation}
\frac{1}{(2\pi)^N
N!}|\Delta(e^{i\delta_1},\ldots,e^{i\delta_N})|^2=
\frac{1}{(2\pi)^N N!}\prod_{j<k}|e^{i\delta_k}-e^{i\delta_j}|^2.
\end{equation}

In random matrix theory it has long been conjectured that in the
limit $N \rightarrow \infty$ the local correlations of the
eigenvalues of random matrices depend exclusively on the
invariance properties of the probability distribution that defines
the ensemble and not on the explicit form of the measure itself.
This random matrix hypothesis is one of the most important
features of the subject. Mathematically it translates into the
statement that, provided that the local eigenvalue density has the
same asymptotic behaviour, the local correlations are mainly
determined by the absolute value of (powers) of the Vandermonde,
whose origin is essentially geometrical. Therefore, we conjecture
that in the limit $N\to\infty$ the local statistics of the roots of the polynomials in
$\mathcal{C}_N$ will be \emph{independent} of our choice of the
joint probability density function for the coefficients of the
polynomials in $\mathcal{R}_N$. Indeed, because of this reason the
extra factor $|e_{N/2}|$ appearing in equation~\eqref{evencasef}
of theorem~\ref{thm:complexcase} when $N$ is even should not
affect the local correlations of the roots in the limit $N
\rightarrow \infty$.

Bogomolny \emph{et al.}~\cite{kn:bbl96} studied self-reciprocal
polynomials  whose coefficients are independent complex normal
random variables and computed the two-point correlation function
$R_2(\delta)$ of the subset of zeros that lie on the unit circle.
As $\delta \rightarrow 0$ they observed linear repulsion between
the arguments of the zeros.  Such level repulsion is a direct
consequence of the Vandermonde that appears in
equations~\eqref{oddcasef} and~\eqref{evencasef} of
theorem~\ref{thm:complexcase} (or more appropriately for their
case in equations~\eqref{oddcase2f} and~\eqref{evencase2f} of
lemma~\ref{lem:cgeneral}) and supports our conjecture.

The appearance of CUE statistics for arithmetic $L$-functions has
been compared to  the appearance of CUE statistics in a chaotic
system without time-reversal
symmetry~\cite{kn:keating93,kn:berrykeating99}. Indeed, the
appearance of the CUE statistics for zeros of $L$-functions has
been heuristically explained by the analogy between the periodic
orbit sum for the density of states of a classically chaotic
system with no time-reversal symmetry and the density of zeros of
the Riemann zeta function expressed as a sum over primes.  Our
observation on the effect of the Euler product can be viewed as
further evidence for that point of view.

In their article Bogomolny \emph{et al.}~\cite{kn:bbl96} commented
that the linear behaviour of $R_2(\delta)$ in the limit $\delta
\rightarrow 0$ was surprising, because it is typical of quantum
mechanical systems whose dynamics is invariant under time
reversal.  Instead, they would have expected quadratic repulsion
between the roots lying on the circle, which is characterizes
chaotic systems whose dynamics is not time reversal invariant and
that self-reciprocal polynomials with arbitrary complex
coefficients were expected to model.  In the case that we study we
are facing the same paradox.  We are modelling \emph{L}-functions
having a functional equation but no Euler product and observe that
the zeros lying on the critical line should have the same local
correlations as the eigenvalues of matrices in the COE ensemble.
Instead, the zeros on the critical line of \emph{L}-functions that
have the further constraint of being expressed in terms of an
Euler product are correlated like the eigenvalues of matrices in
the CUE ensemble. The natural expectation would be the opposite,
since matrices in the CUE ensemble are not in any way restricted,
except that the measure on their set should be invariant under
group multiplication.  The COE ensemble is obtained by imposing
extra symmetry constraints on the matrices in the CUE, whose set,
therefore, includes the set of matrices forming the COE.

 We also consider the case of real self-reciprocal
polynomials. That is, polynomials of the form
\eqref{eqn:defofpoly} satisfying~\eqref{eqn:selfrecip}, where the
$a_n$ are real.  These polynomials have their zeros in complex
conjugate pairs,  and in particular the zeros near $z=1$ would be
expected to have somewhat different behavior than the bulk of the
zeros.   Anomalous behavior of the low-lying zeros occurs for
families of arithmetic $L$-functions, and this behavior has been
modelled by the low-lying eigenvalues of matrices from the
classical Symplectic and Orthogonal groups.

\begin{theorem} \label{thm:realcase}  Let $N=2M$.  Consider
random real monic polynomials $z^{N} + \sum_{n=1}^{N} a_n z^{N-n}$
satisfying the self-reciprocal property $a_{n}={a}_{N-n}$, with
$a_n$ chosen independently and uniformly in the interval $
|a_n|\le \binom{N}{n}$, and restrict to those polynomials having
all zeros on the unit circle.  The joint probability density
function of the set of zeros
$e^{it_1},e^{-it_1},\ldots,e^{it_M},e^{-it_M}$ is given, up to a
normalization constant, by
\begin{equation}
\prod_m|e^{i t_m} - e^{-i t_m}| \;\prod_{j<k}
|e^{it_k}-e^{it_j}|\;|e^{it_k}-e^{-it_j}|.
\end{equation}
In the case where the polynomial has odd degree $N=2M+1$ with
roots $-1$, $e^{it_1},e^{-it_1},$ \ldots, $e^{it_M},e^{-it_M}$
then the joint probability density function of
$e^{it_1},e^{-it_1},\ldots,e^{it_M},e^{-it_M}$ is the same as
given above.
\end{theorem}

The above measure can be
written as, up to a normalization constant,
\begin{equation}
\bigg|\prod_m \sin(t_m) \prod_{j<k}
 \sin\(\frac{t_k-t_j}{2}\) \sin\(\frac{t_k+t_j}{2}\)\bigg|.
\end{equation}
We note that this is the \emph{square root} of Haar measure of
$USp(N)$.  Thus, random real self-reciprocal polynomials,
restricted to have all their zeros on the unit circle, do show
anomalous spacings in their low lying zeros.  But it is not the
same anomalous spacing that has previously been found in families
of arithmetic $L$-functions.  This suggests that for the low-lying
zeros of a family of $L$-functions with real coefficients, the
Euler product has an effect on the vertical spacing of the zeros,
and those $L$-functions behave differently than random real
Dirichlet series with functional equation which just happen to
have their first few zeros on the critical line.

In the next section we prove the Theorems and in the following
sections we give the Jacobian calculations required in the proofs.
We thank Christopher Sinclair for helpful information.

\section{Proofs of the theorems}

We are given a measure on a set of polynomials described in terms
of the coefficients of the polynomial, and we wish to describe the
measure in terms of the roots of the polynomial. Therefore we must
compute the Jacobian of the change of variables from the
coefficients to the roots.  It is well known that the Jacobian is
the Vandermonde in the roots in the case of polynomials with real
coefficients, and the Vandermonde squared in the case of complex
coefficients. Thus we expect the Jacobian to be close to a
Vandermonde in the case of self-reciprocal polynomials, but we
were unable to find all the results we needed in the literature,
so we give a complete proof below.

If $X$ is a random variable then we let
$\langle X \rangle$ denote the expected value of~$X$.
In our case $f$ will be a random polynomial and $X=M[f]$ is
some function of $f$, and we will need to compute
$\langle  M[f] \rangle$.

In Theorem~\ref{thm:complexcase}, consider $N$ even so
$a_1,\ldots,a_{N/2}$ determine~$f$. If
$\rho_{\text{coeffs}}(a_1,\ldots,a_{N/2})$ is a probability
measure on the coefficients of $f$, which is supported on the set
$S$, then
\begin{equation}
\langle M[f]\rangle = \int\limits_{S} M[f]
\rho_{\text{coeffs}}(a_1,\dots,a_{{N/2}}) da_1 \cdots da_{{N/2}} ,
\end{equation}
and by definition $\int\limits_{S}
\rho_{\text{coeffs}}(a_1,\dots,a_{{N/2}}) da_1 \cdots
da_{{N/2}}=1$.

We can express $\langle  M[f] \rangle$ in terms of the  zeros
$z_1,\ldots,z_{N}$
 of~$f$.  Set
\begin{equation}
\rho_{\text{zeros}}(z_1,\ldots,z_{N})
=
\rho_{\text{coeffs}}(a_1(z_1,\ldots,z_{N}),\dots,a_{N}(z_1,\ldots,z_{N}))
\end{equation}
and let $J_\C(z_1,\ldots,z_{N})$ be the Jacobian of the
transformation from the coefficients to the zeros.
Then
\begin{equation}
\langle M[f]\rangle = \int\limits_{S'} M[f]
\rho_{\text{zeros}}(z_1,\dots,z_{N}) |J_\C(z_1,\ldots,z_N)| dz_1
\cdots dz_{N} ,
\end{equation}
where $S'$ is the image of $S$ under the coordinate change.
To prove the Theorems we merely specialize this discussion
to our particular cases.

\begin{proof}[Proof of Theorem \ref{thm:complexcase}]
We have that $\rho_{\text{coeffs}}$ is constant and $S$ is the set
of coefficients of self-reciprocal polynomials having all their
zeros on the unit circle.  So~$\rho_{\text{zeros}}$ is also
constant and $S'=S^1\times\cdots\times S^1$ where $S^1$ is the
unit circle. Thus, we only require the Jacobian of the
transformation, which is given in the following Lemma.

\begin{lemma}\label{lem:complex} If the self-reciprocal polynomial
$f(z)=z^N+\sum_{n=1}^N a_n z^{N-n} $ has all its zeros on the unit
circle, $e^{i\delta_1},\ldots,e^{i\delta_N}$,  then when $N$ is
odd, the absolute value of the Jacobian of the transformation from
coefficient variables $\Re a_1,\Im a_1,\ldots,\Re a_{(N-1)/2},\Im
a_{(N-1)/2},\phi$ (where $a_N=e^{i\phi}$) to the zero variables
$\delta_1,\ldots,\delta_N$ is given by
\begin{equation}
\big|J_\C(e^{i\delta_1},\ldots,e^{i\delta_N})\big|  =
2^{-(N-1)/2}\left|
\Delta(e^{i\delta_1},\ldots,e^{i\delta_N})\right|.
\end{equation}

When $N$ is even, the absolute value of the Jacobian of the
transformation from coefficients $\Re a_1,\Im a_1,\ldots,\Re
a_{N/2},\Im a_{N/2}$ to zeros $\delta_1,\ldots,\delta_N$ is given
by
\begin{equation}
\big|J_\C(e^{i\delta_1},\ldots,e^{i\delta_N})\big|  = 2^{-\frac
N2} \left| e_{N/2}(e^{i\delta_1},\ldots,e^{i\delta_N})\;
\Delta(e^{i\delta_1},\ldots,e^{i\delta_N})\right|,
\end{equation}
where $e_{N/2}(e^{i\delta_1},\ldots,e^{i\delta_N})$ is the $N/2$
elementary symmetric function.

\end{lemma}
A proof of the Lemma can be found in~Section~\ref{sec:clemma}.

Assembling the pieces we have, for example for odd $N$,
\begin{equation}
\langle M[f]\rangle = \frac{1}{(4\sqrt{\pi})^N\Gamma(1 +
N/2)}\int_0^{2\pi}\cdots \int_0^{2\pi} M[f]
|\Delta(e^{i\delta_1},\ldots,e^{i\delta_N})| d\delta_1 \cdots
d\delta_{N} ,
\end{equation}
which is equivalent to Theorem~\ref{thm:complexcase}.
\end{proof}

The proof of Theorem~\ref{thm:realcase} is identical except that
we require the following lemma, which is proven in Section~\ref{sec:rlemma}.

\begin{lemma}\label{lem:real} If the real self-reciprocal polynomial
$f$ has even degree $N=2M$ and has all its zeros on the unit
circle, then the absolute value of the Jacobian of the
transformation from coefficients $a_1,\ldots,a_{M}$ to zeros
$e^{it_1},e^{-it_1},\ldots,e^{it_M},e^{-it_{M}}$ is given by
\begin{equation}
| J_\R(e^{it_1},e^{-it_1},\ldots,e^{it_M},e^{-it_{M}}) | =
\left| \prod_m(e^{it_m} - e^{-it_m}) \prod_{j<k}
(e^{it_k}-e^{it_j})(e^{it_k}-e^{-it_j}) \right| .
\end{equation}
In the case the degree $N=2M+1$ is odd, and $f$ has zeros at $-1$,
$e^{it_1},e^{-it_1},\ldots,e^{it_M},e^{-it_{M}}$, the Jacobian is
again given by the above formula.
\end{lemma}

\section{Calculation of the Jacobian: complex case}\label{sec:clemma}

We prove the following generalization of Lemma~\ref{lem:complex}.

\begin{lemma}
\label{lem:cgeneral}
Let the roots of a self-reciprocal polynomial $f$ be
$\alpha_1=e^{i\delta_1},\ldots,\alpha_L=e^{i\delta_L}$, for those
roots on the unit circle,
 and
$\beta_1=\rho_1e^{i\theta_1},\tfrac{1}{\overline{\beta}_1}=\tfrac{e^{i\theta_1}}{\rho_1},
\ldots,\beta_M=\rho_Me^{i\theta_M},\tfrac{1}{\overline{\beta}_M}=\tfrac{e^{i\theta_M}}{\rho_M}$
for the roots occurring in pairs off the unit circle.  When
$N=L+2M$ is odd, the absolute value of the Jacobian of the
transformation from coefficients $\Re a_1,\Im a_1,\ldots,\Re
a_{(N-1)/2},\Im a_{(N-1)/2},\phi$ (where $a_N=e^{i\phi}$) to zeros
$\rho_1,\theta_1,\rho_2,\theta_2$, $\ldots\,$, $
\rho_M,\theta_M,\delta_1,\ldots,\delta_L$ is given by
\begin{multline}
\label{oddcase2f}
\big|J_\C(\rho_1,\theta_1,\rho_2,\theta_2$,
$\ldots\,$, $ \rho_M,\theta_M,\delta_1,\ldots,\delta_L)\big|
\\ =2^{M - (N-1)/2}\;\left| \left(\prod_{m=1}^M
\frac{1}{\rho_m}\right) \;
\Delta(\beta_1,\tfrac{1}{\overline{\beta}_1},\beta_2,\tfrac{1}{\overline{\beta}_2},\ldots,
\beta_m,\tfrac{1}{\overline{\beta}_m},\alpha_1,\ldots,\alpha_L)\right|.
\end{multline}

When $N=L+2M$ is even, the absolute value of the Jacobian of the
transformation from coefficients $\Re a_1,\Im a_1,\ldots,\Re
a_{N/2},\Im a_{N/2}$ to zeros $\rho_1,\theta_1,\rho_2,\theta_2$,
$\ldots\,$, $ \rho_M,\theta_M,\delta_1,\ldots,\delta_L$ is given
by
\begin{equation}
\begin{split}
\label{evencase2f} \big|J_\C(\rho_1,\theta_1,&\rho_2,\theta_2,
\ldots\,, \rho_M,\theta_M,\delta_1,\ldots,\delta_L)\big|
\\ & = 2^{M - \frac N2}\Biggl|\left(\prod_{m=1}^M
\frac{1}{\rho_m}\right) \;
e_{N/2}(\beta_1,\tfrac{1}{\overline{\beta}_1},\beta_2,\tfrac{1}{\overline{\beta}_2},\ldots,
\beta_m,\tfrac{1}{\overline{\beta}_m},\alpha_1,\ldots,\alpha_L)\;\\
& \quad \times
\Delta(\beta_1,\tfrac{1}{\overline{\beta}_1},\beta_2,\tfrac{1}{\overline{\beta}_2},\ldots,
\beta_m,\tfrac{1}{\overline{\beta}_m},\alpha_1,\ldots,\alpha_L)\Bigr|,
\end{split}
\end{equation}
where the $e_{N/2}$ is the $N/2$ elementary symmetric function.
\end{lemma}

\begin{proof}
The polynomial $f(z)$ has order $N=L+2M$:
\begin{equation}
\begin{split}
f(z)&=(z-\alpha_1)(z-\alpha_2)\cdots(z-\alpha_L)(z-\beta_1)(z-\tfrac{1}{\overline{\beta}_1})
\cdots(z-\beta_M)(z-\tfrac{1}{\overline{\beta}_M})\\
&=z^N+a_1z^{N-1}+a_2z^{N-2}+\cdots+a_{N-2}z^2+a_{N-1}z+a_N.
\end{split}
\end{equation}
Note that if we define
\begin{subequations}\label{eq:alphabeta}
\begin{align}
\alpha_j & =e^{i\delta_j},  &&  j =1,\ldots,L\\
\label{beq}
\beta_j & =t_j\rho_j=e^{i\theta_j}\rho_j,  && j =1,\ldots,M\\
\frac{1}{\overline{\beta}_j}& \label{neq}
=\frac{t_j}{\rho_j}=\frac{e^{i\theta_j}}{\rho_j}, &&  j
=1,\ldots,M
\end{align}
\end{subequations}
(with $\delta_j$, $\rho_j$ and $\theta_j$ real) then $a_N$ is on
the unit circle and
\begin{equation}
a_N=(-1)^Ne^{2i\theta_1}e^{2i\theta_2}\cdots e^{2i\theta_M}
e^{i\delta_1}\cdots e^{i\delta_L}=e^{i\phi}.
\end{equation}
Also
\begin{equation}
a_j=(-1)^je_j(\alpha_1,\ldots,\alpha_L,\beta_1,\tfrac{1}{\overline{\beta}_1},
\ldots,\beta_M,\tfrac{1}{\overline{\beta}_M}),
\end{equation}
where $e_j(x_1,\ldots,x_n)$ is the $j$th elementary symmetric
function.  In addition we have
\begin{equation}
a_{N-j}=a_N\overline{a}_j,
\end{equation}
since by construction $f$ is self-reciprocal.

For now we take $N$ odd; the slight variation when $N$ is even is
described at the end of this section.  We want the Jacobian of the
transformation from the independent real variables $\Re a_1, \Im
a_1,\ldots,\Re a_{\frac{N-1}{2}},\Im a_{\frac{N-1}{2}},\phi$ to
the real independent variables $\rho_1,\theta_1,\rho_2,\theta_2$,
$\ldots\,$, $ \rho_M,\theta_M,\delta_1,\ldots,\delta_L$.

So, the Jacobian is:
\begin{equation}
\begin{split}
J_\C& = \begin{vmatrix} \frac{\partial \Re a_1}{\partial \rho_1}
&\frac{\partial \Re  a_1}{\partial \theta_1} & \cdots
&\frac{\partial \Re  a_1}{\partial \rho_M} & \frac{\partial \Re
a_1}{\partial \theta_M} & \frac{\partial \Re a_1}{\partial
\delta_1} &\cdots & \frac{\partial \Re
a_1}{\partial \delta_L} \\
\frac{\partial \Im  a_1}{\partial \rho_1} &\frac{\partial \Im
a_1}{\partial \theta_1} & \cdots &\frac{\partial \Im a_1}{\partial
\rho_M} & \frac{\partial \Im  a_1}{\partial \theta_M} &
\frac{\partial \Im  a_1}{\partial \delta_1} &\cdots &
\frac{\partial \Im
a_1}{\partial \delta_L} \\
\vdots&\vdots & \ddots & \vdots & \vdots  & \vdots & \ddots
&\vdots \\
\frac{\partial \Re  a_{\frac{N-1}{2}}}{\partial \rho_1}
&\frac{\partial \Re  a_{\frac{N-1}{2}}}{\partial \theta_1} &
\cdots &\frac{\partial \Re  a_{\frac{N-1}{2}}}{\partial \rho_M} &
\frac{\partial \Re  a_{\frac{N-1}{2}}}{\partial \theta_M} &
\frac{\partial \Re  a_{\frac{N-1}{2}}}{\partial \delta_1} &\cdots
& \frac{\partial \Re
a_{\frac{N-1}{2}}}{\partial \delta_L} \\
\frac{\partial \Im  a_{\frac{N-1}{2}}}{\partial \rho_1}
&\frac{\partial \Im  a_{\frac{N-1}{2}}}{\partial \theta_1} &
\cdots &\frac{\partial \Im  a_{\frac{N-1}{2}}}{\partial \rho_M} &
\frac{\partial \Im  a_{\frac{N-1}{2}}}{\partial \theta_M} &
\frac{\partial \Im  a_{\frac{N-1}{2}}}{\partial \delta_1} &\cdots
& \frac{\partial \Im
a_{\frac{N-1}{2}}}{\partial \delta_L} \\
 \frac{\partial \phi}{\partial \rho_1} &\frac{\partial \phi}{\partial
\theta_1} & \cdots &\frac{\partial \phi}{\partial \rho_M} &
\frac{\partial \phi}{\partial \theta_M} & \frac{\partial
\phi}{\partial \delta_1} &\cdots & \frac{\partial \phi}{\partial
\delta_L}\end{vmatrix}\\
 &=(-\tfrac{1}{2})^{\frac{N-1}{2}}
\begin{vmatrix} \frac{\partial  a_1}{\partial \rho_1}
&\frac{\partial a_1}{\partial \theta_1} & \cdots &\frac{\partial
a_1}{\partial \rho_M} & \frac{\partial  a_1}{\partial \theta_M} &
\frac{\partial
 a_1}{\partial \delta_1} &\cdots & \frac{\partial
a_1}{\partial \delta_L} \\&&&&&&&\\ \frac{\partial
\overline{a}_1}{\partial \rho_1} &\frac{\partial
\overline{a}_1}{\partial \theta_1} & \cdots &\frac{\partial
\overline{a}_1}{\partial \rho_M} & \frac{\partial
\overline{a}_1}{\partial \theta_M} & \frac{\partial
 \overline{a}_1}{\partial \delta_1} &\cdots & \frac{\partial
\overline{a}_1}{\partial \delta_L} \\ \vdots&\vdots & \ddots &
\vdots & \vdots  & \vdots & \ddots
&\vdots \\
\frac{\partial  a_{\frac{N-1}{2}}}{\partial \rho_1}
&\frac{\partial  a_{\frac{N-1}{2}}}{\partial \theta_1} & \cdots
&\frac{\partial  a_{\frac{N-1}{2}}}{\partial \rho_M} &
\frac{\partial  a_{\frac{N-1}{2}}}{\partial \theta_M} &
\frac{\partial  a_{\frac{N-1}{2}}}{\partial \delta_1} &\cdots &
\frac{\partial a_{\frac{N-1}{2}}}{\partial \delta_L} \\&&&&&&&\\
\frac{\partial \overline{ a}_{\frac{N-1}{2}}}{\partial \rho_1}
&\frac{\partial \overline{a}_{\frac{N-1}{2}}}{\partial \theta_1} &
\cdots &\frac{\partial \overline{a}_{\frac{N-1}{2}}}{\partial
\rho_M} & \frac{\partial \overline{a}_{\frac{N-1}{2}}}{\partial
\theta_M} & \frac{\partial \overline{ a}_{\frac{N-1}{2}}}{\partial
\delta_1} &\cdots & \frac{\partial \overline{
a}_{\frac{N-1}{2}}}{\partial \delta_L} \\&&&&&&&\\
 \frac{\partial \phi}{\partial \rho_1} &\frac{\partial \phi}{\partial
\theta_1} & \cdots &\frac{\partial \phi}{\partial \rho_M} &
\frac{\partial \phi}{\partial \theta_M} & \frac{\partial
\phi}{\partial \delta_1} &\cdots & \frac{\partial \phi}{\partial
\delta_L}\end{vmatrix}.
\end{split}
\end{equation}

This step was achieved in two stages: first, by adding each even
row to the one above, and then by multiplying each even row by $-1$
and adding to it $1/2$ the row above.

Now note that
\begin{subequations}
\begin{align}
\frac{\partial \phi}{\partial x} &= \frac{\partial\phi} {\partial
a_N} \frac{\partial a_N}{\partial x} = \frac{\partial
a_N}{\partial x} \left( \frac{1}{ia_N}\right),\\
\frac{\partial a_{N-j}} {\partial x} &= a_N \frac{\partial
\overline{a}_j}{\partial x} + \overline{a}_j \frac{\partial
a_N}{\partial x}
\end{align}
\end{subequations}
So we have
\begin{equation}
\begin{split}
J_\C=\left(-\frac{1}{2}\right)^{\frac{N-1}{2}} & \left(
\frac{1}{ia_N}\right) \left( \frac{1}{a_N}\right)^{\frac{N-1}{2}}
\\
&\times \begin{vmatrix} \frac{\partial a_1}{\partial \rho_1}
&\frac{\partial a_1}{\partial \theta_1} & \cdots &\frac{\partial
a_1}{\partial \rho_M} & \frac{\partial a_1}{\partial \theta_M} &
\frac{\partial
 a_1}{\partial \delta_1} &\cdots & \frac{\partial
a_1}{\partial \delta_L} \\
\frac{\partial  a_{N-1}}{\partial \rho_1} &\frac{\partial
a_{N-1}}{\partial \theta_1} & \cdots &\frac{\partial
a_{N-1}}{\partial \rho_M} & \frac{\partial a_{N-1}}{\partial
\theta_M} & \frac{\partial
 a_{N-1}}{\partial \delta_1} &\cdots & \frac{\partial
a_{N-1}}{\partial \delta_L} \\
\vdots&\vdots & \ddots & \vdots & \vdots  & \vdots & \ddots
&\vdots \\
\frac{\partial  a_{\frac{N-1}{2}}}{\partial \rho_1}
&\frac{\partial  a_{\frac{N-1}{2}}}{\partial \theta_1} & \cdots
&\frac{\partial  a_{\frac{N-1}{2}}}{\partial \rho_M} &
\frac{\partial  a_{\frac{N-1}{2}}}{\partial \theta_M} &
\frac{\partial a_{\frac{N-1}{2}}}{\partial \delta_1} &\cdots &
\frac{\partial
a_{\frac{N-1}{2}}}{\partial \delta_L} \\
\frac{\partial  a_{\frac{N+1}{2}}}{\partial \rho_1}
&\frac{\partial a_{\frac{N+1}{2}}}{\partial \theta_1} & \cdots
&\frac{\partial a_{\frac{N+1}{2}}}{\partial \rho_M} &
\frac{\partial a_{\frac{N+1}{2}}}{\partial \theta_M} &
\frac{\partial a_{\frac{N+1}{2}}}{\partial \delta_1} &\cdots &
\frac{\partial
a_{\frac{N+1}{2}}}{\partial \delta_L} \\
 \frac{\partial a_N}{\partial \rho_1} &\frac{\partial a_N}{\partial
\theta_1} & \cdots &\frac{\partial a_N}{\partial \rho_M} &
\frac{\partial a_N}{\partial \theta_M} & \frac{\partial
a_N}{\partial \delta_1} &\cdots & \frac{\partial a_N}{\partial
\delta_L}\end{vmatrix}.
\end{split}
\end{equation}
So we have, where $\varepsilon$ is a quantity with modulus one that
may vary at each occurrence,
\begin{equation}\label{eqn:rhoanda}
J_\C= \varepsilon\left( \frac{1}{2}\right)^{\frac{N-1}{2}} \left|
\begin{array}{cccccccc} \frac{\partial a_1}{\partial \rho_1} &
\frac{\partial a_1}{\partial \theta_1} & \cdots& \frac{\partial
a_1}{\partial \rho_M} & \frac{\partial a_1}{\partial \theta_M} &
\frac{\partial a_1}{\partial \delta_1}&\cdots& \frac{\partial
a_1}{\partial \delta_L}\\
\vdots & \vdots &\ddots & \vdots &\vdots&\vdots &\ddots & \vdots\\
\frac{\partial a_N}{\partial \rho_1} & \frac{\partial
a_N}{\partial \theta_1} & \cdots& \frac{\partial a_N}{\partial
\rho_M} & \frac{\partial a_N}{\partial \theta_M} & \frac{\partial
a_N}{\partial \delta_1}&\cdots& \frac{\partial a_N}{\partial
\delta_L}\end{array}\right|.
\end{equation}

For convenience let us set $\nu_j=1/\overline{\beta_j}$. (We thank
the editor for suggesting how to improve the next part of the
proof. Our argument was more lengthy.) The proof of the lemma is
achieved by expressing the determinant~\eqref{eqn:rhoanda} in
terms of
\begin{multline}\label{eq:jaco1}
\left| \begin{array}{cccccccc}\frac{\partial a_1}{\partial
\beta_1} & \frac{\partial a_1}{\partial \nu_1} & \cdots&
\frac{\partial a_1}{\partial \beta_M} & \frac{\partial
a_1}{\partial \nu_M} & \frac{\partial a_1}{\partial
\alpha_1}&\cdots& \frac{\partial
a_1}{\partial \alpha_L}\\
\vdots & \vdots &\ddots & \vdots &\vdots&\vdots &\ddots & \vdots\\
\frac{\partial a_N}{\partial \beta_1} & \frac{\partial
a_N}{\partial \nu_1} & \cdots& \frac{\partial a_N}{\partial
\beta_M} & \frac{\partial a_N}{\partial \nu_M} & \frac{\partial
a_N}{\partial \alpha_1}&\cdots& \frac{\partial a_N}{\partial
\alpha_L}\end{array}\right|\\=(-1)^N\;
\Delta(\beta_1,\nu_1,\ldots,\beta_M,\nu_M,\alpha_1,\ldots,\alpha_L).
\end{multline}
To prove this equality note that the left hand side is a
homogeneous polynomial of order $N(N-1)/2$ in the variables
$\beta_1,\nu_1,\ldots,$ $\beta_M,\nu_M,\alpha_1,\ldots,\alpha_L$.
It is also antisymmetric under exchange of any two of the
variables. Therefore, up to a constant, it must be proportional to
the Vandermonde $\Delta(\beta_1,\nu_1,\ldots,$
$\beta_M,\nu_M,\alpha_1,\ldots,\alpha_L)$ (defined in
(\ref{eq:vand})).  The constant $(-1)^N$ can be determined as
follows.  Let us denote by $x_1,\ldots,x_N$ the  variables
$\beta_1,\nu_1,\ldots,$ $\beta_M,\nu_M,\alpha_1,\ldots,\alpha_L$.
The generic element in the Jacobian~\eqref{eq:jaco1} is
\begin{equation}
\label{Jel}
\frac{\partial a_j}{\partial x_k} = (-1)^j e_{j-1,k},
\end{equation}
where $e_{j,k}$ is the $j$-th symmetric function in the $x_l$
excluding the variable $x_k$.  We want to compute the coefficient
in front of
\begin{equation}
\label{prodx}
x_1^{N-1}x_2^{N-2} \cdots x_N.
\end{equation}
Now, any $x_j$ in any element of the Jacobian matrix appears at
most with exponent one. Furthermore, for any given choice of $j$
indices $l_1,l_2,\ldots,l_j$, ($l_m \neq k$)  the product
\begin{equation}
x_{l_1}x_{l_2}\cdots x_{l_j}
\end{equation}
appears once and only once in the symmetric function $e_{j,k}$. It
follows that the monomial~\eqref{prodx} can only come from the
term
\begin{equation}
\frac{\partial a_1}{\partial x_1}\frac{\partial a_2}{\partial x_2}
\cdots \frac{\partial a_N}{\partial x_N}
\end{equation}
in the expansion of the Jacobian and that the coefficient in front
of it is $(-1)^{N(N+1)/2}$.  The constant $(-1)^N$ is obtained by
observing that the monomial~\eqref{prodx} appear with a factor
$(-1)^{N(N-1)/2}$ in the expansion of the Vandermonde.

The relation between the determinants~\eqref{eqn:rhoanda}
and~\eqref{eq:jaco1} is obtained by analytically continuing
equations~\eqref{eq:alphabeta} in the complex planes of $\rho$ and
$\theta$ and then by changing the variables from
$\rho_1,\theta_1,\ldots,\rho_M,\theta_M,\delta_1,\ldots,\delta_L$
to $\beta_1,\nu_1,\ldots,\beta_M,\nu_M,\alpha_1,\ldots,\alpha_L$.
Thus, we incur the further Jacobian
\begin{equation}\label{eq:jaco2}
\left|\begin{array}{cccccccc} \frac{\partial \rho_1}{\partial
\beta_1} & \frac{\partial \rho_1}{\partial \nu_1} & \cdots&
\frac{\partial \rho_1}{\partial \beta_M} & \frac{\partial
\rho_1}{\partial \nu_M} & \frac{\partial \rho_1}{\partial
\alpha_1}&\cdots& \frac{\partial \rho_1}{\partial
\alpha_L}\\
\frac{\partial \theta_1}{\partial \beta_1} & \frac{\partial
\theta_1}{\partial \nu_1} & \cdots& \frac{\partial
\theta_1}{\partial \beta_M} & \frac{\partial \theta_1}{\partial
\nu_M} & \frac{\partial \theta_1}{\partial \alpha_1}&\cdots&
\frac{\partial
\theta_1}{\partial \alpha_L}\\
\vdots & \vdots &\ddots & \vdots &\vdots&\vdots &\ddots & \vdots\\
\frac{\partial \rho_M}{\partial \beta_1} & \frac{\partial
\rho_M}{\partial \nu_1} & \cdots& \frac{\partial \rho_M}{\partial
\beta_M} & \frac{\partial \rho_M}{\partial \nu_M} & \frac{\partial
\rho_M}{\partial \alpha_1}&\cdots& \frac{\partial \rho_M}{\partial
\alpha_L}\\
\frac{\partial \theta_M}{\partial \beta_1} & \frac{\partial
\theta_M}{\partial \nu_1} & \cdots& \frac{\partial
\theta_M}{\partial \beta_M} & \frac{\partial \theta_M}{\partial
\nu_M} & \frac{\partial \theta_M}{\partial \alpha_1}&\cdots&
\frac{\partial
\theta_M}{\partial \alpha_L}\\
\frac{\partial \delta_1}{\partial \beta_1} & \frac{\partial
\delta_1}{\partial \nu_1} & \cdots& \frac{\partial
\delta_1}{\partial \beta_M} & \frac{\partial \delta_1}{\partial
\nu_M} & \frac{\partial \delta_1}{\partial \alpha_1}&\cdots&
\frac{\partial \delta_1}{\partial
\alpha_L}\\\vdots & \vdots &\ddots & \vdots &\vdots&\vdots &\ddots & \vdots\\
\frac{\partial \delta_L}{\partial \beta_1} & \frac{\partial
\delta_L}{\partial \nu_1} & \cdots& \frac{\partial
\delta_L}{\partial \beta_M} & \frac{\partial \delta_L}{\partial
\nu_M} & \frac{\partial \delta_L}{\partial \alpha_1}&\cdots&
\frac{\partial \delta_L}{\partial \alpha_L}\end{array}\right|.
\end{equation}
Using equations~(\ref{eq:alphabeta}) the computation of this
determinant is straightforward. The matrix is all zero except for
$2\times 2$ blocks on the diagonal for the first $2M$ columns, and
then single elements on the diagonal for the last $L$ columns. A
given $2\times 2$ block looks like
\begin{equation}
\left|\begin{array}{cc}\frac{\partial \rho_j}{\partial \beta_j} &
\frac{\partial \rho_j}{\partial \nu_j}\\ \frac{\partial
\theta_j}{\partial \beta_j} & \frac{\partial \theta_j}{\partial
\nu_j}\end{array}\right|=\left|\begin{array}{cc}
\frac{1}{2\sqrt{\beta_j \nu_j}} &
-\frac{1}{2\nu_j}\sqrt{\frac{\beta_j}{\nu_j}}\\
-\frac{i}{2\beta_j} &
-\frac{i}{2\nu_j}\end{array}\right|=-\frac{i}{2\nu_j\sqrt{\beta_j\nu_j}},
\end{equation}
and the single elements on the diagonal have modulus one.  So,
dividing (\ref{eq:jaco1}) by (\ref{eq:jaco2}) we have
\begin{equation}
J_\C= \varepsilon\;2^{M- (N-1)/2} \;\left(\prod_{m=1}^M
\frac{1}{\rho_m}\right) \;
\Delta(\beta_1,\tfrac{1}{\overline{\beta}_1},\beta_2,\tfrac{1}{\overline{\beta}_2},\ldots,
\beta_m,\tfrac{1}{\overline{\beta}_m},\alpha_1,\ldots,\alpha_L).
\end{equation}
Here $\varepsilon$ incorporates quantities that
may depend on 
$\beta_1,\tfrac{1}{\overline{\beta}_1},\beta_2,\tfrac{1}{\overline{\beta}_2},\ldots,
\beta_m,\tfrac{1}{\overline{\beta}_m},\alpha_1,\ldots,\alpha_L$
but has absolute value one.  This completes the proof in the case
$N$ is odd.

When $N$ is even, we want the Jacobian of the transformation from
the real, independent variables $\Re a_1, \Im a_1,\ldots,\Re
a_{\frac{N}{2}},\Im a_{\frac{N}{2}}$ to the real independent
variables $\rho_1,\theta_1,\rho_2,\theta_2$, $\ldots\,$, $
\rho_M,\theta_M,\delta_1,\ldots,\delta_L$.

So, we start with
\begin{equation}\label{eq:Nevencomp}
\begin{split}
J_\C& = \begin{vmatrix} \frac{\partial \Re a_1}{\partial \rho_1}
&\frac{\partial \Re  a_1}{\partial \theta_1} & \cdots
&\frac{\partial \Re  a_1}{\partial \rho_M} & \frac{\partial \Re
a_1}{\partial \theta_M} & \frac{\partial \Re a_1}{\partial
\delta_1} &\cdots & \frac{\partial \Re
a_1}{\partial \delta_L} \\
\frac{\partial \Im  a_1}{\partial \rho_1} &\frac{\partial \Im
a_1}{\partial \theta_1} & \cdots &\frac{\partial \Im a_1}{\partial
\rho_M} & \frac{\partial \Im  a_1}{\partial \theta_M} &
\frac{\partial \Im  a_1}{\partial \delta_1} &\cdots &
\frac{\partial \Im
a_1}{\partial \delta_L} \\
\vdots&\vdots & \ddots & \vdots & \vdots  & \vdots & \ddots
&\vdots \\
\frac{\partial \Re  a_{\frac{N}{2}}}{\partial \rho_1}
&\frac{\partial \Re  a_{\frac{N}{2}}}{\partial \theta_1} & \cdots
&\frac{\partial \Re  a_{\frac{N}{2}}}{\partial \rho_M} &
\frac{\partial \Re  a_{\frac{N}{2}}}{\partial \theta_M} &
\frac{\partial \Re  a_{\frac{N}{2}}}{\partial \delta_1} &\cdots &
\frac{\partial \Re
a_{\frac{N}{2}}}{\partial \delta_L} \\
\frac{\partial \Im  a_{\frac{N}{2}}}{\partial \rho_1}
&\frac{\partial \Im  a_{\frac{N}{2}}}{\partial \theta_1} & \cdots
&\frac{\partial \Im  a_{\frac{N}{2}}}{\partial \rho_M} &
\frac{\partial \Im  a_{\frac{N}{2}}}{\partial \theta_M} &
\frac{\partial \Im  a_{\frac{N}{2}}}{\partial \delta_1} &\cdots &
\frac{\partial \Im
a_{\frac{N}{2}}}{\partial \delta_L} \end{vmatrix}\\
 &=(-\tfrac{1}{2})^{\frac{N}{2}}
\begin{vmatrix} \frac{\partial  a_1}{\partial \rho_1}
&\frac{\partial a_1}{\partial \theta_1} & \cdots &\frac{\partial
a_1}{\partial \rho_M} & \frac{\partial  a_1}{\partial \theta_M} &
\frac{\partial
 a_1}{\partial \delta_1} &\cdots & \frac{\partial
a_1}{\partial \delta_L} \\&&&&&&&\\ \frac{\partial
\overline{a}_1}{\partial \rho_1} &\frac{\partial
\overline{a}_1}{\partial \theta_1} & \cdots &\frac{\partial
\overline{a}_1}{\partial \rho_M} & \frac{\partial
\overline{a}_1}{\partial \theta_M} & \frac{\partial
 \overline{a}_1}{\partial \delta_1} &\cdots & \frac{\partial
\overline{a}_1}{\partial \delta_L} \\ \vdots&\vdots & \ddots &
\vdots & \vdots  & \vdots & \ddots
&\vdots \\
\frac{\partial  a_{\frac{N}{2}}}{\partial \rho_1} &\frac{\partial
a_{\frac{N}{2}}}{\partial \theta_1} & \cdots &\frac{\partial
a_{\frac{N}{2}}}{\partial \rho_M} & \frac{\partial
a_{\frac{N}{2}}}{\partial \theta_M} & \frac{\partial
a_{\frac{N}{2}}}{\partial \delta_1} &\cdots &
\frac{\partial a_{\frac{N}{2}}}{\partial \delta_L} \\&&&&&&&\\
\frac{\partial \overline{ a}_{\frac{N}{2}}}{\partial \rho_1}
&\frac{\partial \overline{a}_{\frac{N}{2}}}{\partial \theta_1} &
\cdots &\frac{\partial \overline{a}_{\frac{N}{2}}}{\partial
\rho_M} & \frac{\partial \overline{a}_{\frac{N}{2}}}{\partial
\theta_M} & \frac{\partial \overline{ a}_{\frac{N}{2}}}{\partial
\delta_1} &\cdots & \frac{\partial \overline{
a}_{\frac{N}{2}}}{\partial \delta_L} \end{vmatrix}.
\end{split}
\end{equation}

Now we want to transform the derivatives of $\overline{a}_j$ into
derivatives of $a_{N-j}$, with the exception of
$\overline{a}_{N/2}$, which should give us derivatives of $a_N$.
Note that $a_N$ has modulus 1 and
\begin{equation}
\frac{a_{N/2}}{\overline{a}_{N/2}} = a_N.
\end{equation}
Therefore, we have
\begin{eqnarray}
\frac{\partial a_N}{\partial x}&=& \frac{\partial}{\partial x}
\left( \frac{a_{N/2}}{\overline{a}_{N/2}}\right)\\
&=& \frac{1}{\overline{a}_{N/2}} \frac{\partial a_{N/2}} {\partial
x} - \frac{a_{N/2}}{(\overline{a}_{N/2})^2} \frac{\partial
\overline{a}_{N/2}}{\partial x}.
\end{eqnarray}
Thus we need to multiply the last row of (\ref{eq:Nevencomp}) by
$-a_{N/2}/(\overline{a}_{N/2})^2$ and add to it
$1/\overline{a}_{N/2}$ times the row above.   This procedure has
multiplied the determinant by a factor
$-a_{N/2}/(\overline{a}_{N/2})^2=-a_{N}/\overline{a}_{N/2}$.  Note
that $a_N$ has modulus one, but $a_{N/2}$ does not.  Thus,
incorporating factors of modulus one into $\varepsilon$, the Jacobian
is
\begin{equation}
J_\C= \varepsilon \;\overline{a}_{N/2}\left(
\frac{1}{2}\right)^{\frac{N}{2}} \left|
\begin{array}{cccccccc} \frac{\partial a_1}{\partial \rho_1} &
\frac{\partial a_1}{\partial \theta_1} & \cdots& \frac{\partial
a_1}{\partial \rho_M} & \frac{\partial a_1}{\partial \theta_M} &
\frac{\partial a_1}{\partial \delta_1}&\cdots& \frac{\partial
a_1}{\partial \delta_L}\\
\vdots & \vdots &\ddots & \vdots &\vdots&\vdots &\ddots & \vdots\\
\frac{\partial a_N}{\partial \rho_1} & \frac{\partial
a_N}{\partial \theta_1} & \cdots& \frac{\partial a_N}{\partial
\rho_M} & \frac{\partial a_N}{\partial \theta_M} & \frac{\partial
a_N}{\partial \delta_1}&\cdots& \frac{\partial a_N}{\partial
\delta_L}\end{array}\right|.
\end{equation}

It is now just necessary to follow the steps as in the odd $N$
case in order to prove Lemma~\ref{lem:cgeneral}.
\end{proof}

\section{Calculation of the Jacobian: real case}\label{sec:rlemma}

We prove Lemma~\ref{lem:real}.  It would be possible to cover
a more general case, considering the number of zeros on the unit circle, the
number in complex conjugate pairs located symmetrically with respect to
the unit circle, and the number of real zeros.
But our concern here
is with the case that all zeros are on the unit circle so we only
consider the case we require in this paper.
We describe the odd degree case in detail and then discuss the
modifications required for the even degree case.

We have a self-reciprocal polynomial $f(z)$ that has a root at $-1$ and
at $\beta_1,\overline{\beta}_1,
\ldots,\beta_M,{\overline{\beta}_M}$ on the unit circle. Thus,
$\overline{\beta}_j=1/\beta_j$.
The polynomial $f(z)$ has order $N=2M+1$:
\begin{equation}
\begin{split}
f(z)&=(z+1)(z-\beta_1)(z-{\overline{\beta}_1})
\cdots(z-\beta_M)(z-{\overline{\beta}_M})\\
&=z^N+a_1z^{N-1}+a_2z^{N-2}+\cdots+a_{N-2}z^2+a_{N-1}z+1.
\end{split}
\end{equation}
We have the functional equation
\begin{equation}
f(z)=z^N f\(\frac{1}{z}\),
\end{equation}
which implies the symmetry of the coefficients
\begin{equation}
a_n = a_{N-n} .
\end{equation}
So the polynomial is determined by $a_1$,\ldots, $a_M$.
We also have
\begin{equation}
a_n=(-1)^ne_n(-1,\beta_1,\overline{\beta}_1,
\ldots,\beta_M,\overline{\beta}_M),
\end{equation}
where $e_n$ is the $n$th elementary symmetric
function.

We want the Jacobian of the transformation from the
real variables $a_1$,\ldots,$a_M$ to the
real variables $t_1$,\ldots,$t_M$, where $\beta_j=e^{i t_j}$.

We start with
\begin{equation}
J_\R :=
\left| \begin{array}{cccccccc}
\frac{\partial
a_1}{\partial t_1} &\frac{\partial a_1}{\partial
t_2} & \cdots &\frac{\partial a_1}{\partial
t_M}
 \\
\vdots&\vdots & \ddots & \vdots
 \\
\frac{\partial a_{M}}{\partial t_1}
&\frac{\partial a_{M}}{\partial t_2} &
\cdots &\frac{\partial  a_{M}}{\partial
t_M}
 \\
\end{array}\right|
=
i^M
\beta_1\cdots \beta_M
\left| \begin{array}{cccccccc}
\frac{\partial
a_1}{\partial \beta_1} &\frac{\partial a_1}{\partial
\beta_2} & \cdots &\frac{\partial a_1}{\partial
\beta_M}
 \\
\vdots&\vdots & \ddots & \vdots
 \\
\frac{\partial a_{M}}{\partial \beta_1}
&\frac{\partial a_{M}}{\partial \beta_2} &
\cdots &\frac{\partial  a_{M}}{\partial
\beta_M}
 \\
\end{array}\right| ,
\end{equation}
where we used the fact that
\begin{equation}
\frac{\partial a_j}{\partial t_k}
=
\frac{\partial a_j}{\partial \beta_k}
\frac{d \beta_k}{d t_k}
=i \beta_k \frac{\partial a_j}{\partial \beta_k} .
\end{equation}

At this point our calculation becomes somewhat different than the complex
case in the previous section.  It is possible that similar methods could be used here,
but the argument would be more sophisticated because (as we will see), the
answer is not symmetric in all the variables.
We will directly calculate the Jacobian, instead of deducing the form
of the answer from its various symmetries.

Following the method in~\cite{Sinc}, for each $m$ we have
\begin{equation}
a_n=(-1)^n\left( \({\beta_m}+\frac{1}{\beta_m}\) e'_{n-1,m} +
e'_{n-2,m}+e'_{n,m}\right),
\end{equation}
where
we write
$e'_{n,m}$ for the $n$th symmetric
 function in $-1$ and all the
 $\beta_j$ and $\overline{\beta}_j=\frac{1}{{\beta_j}}$, $j=1,\ldots,M$, except
 for $\beta_m$ and $\frac{1}{{\beta_m}}$,
with the convention that
$e_0=e'_{0,m}=1$ and
 $e_n=e'_{n,m}=0$ if $n$ is greater than the number of variables.
Thus, we obtain
 \begin{eqnarray}
 \frac{\partial a_n}{\partial \beta_m} &=& (-1)^n \left(
\(1-\frac{1}{\beta_m^2}\) e'_{n-1,m}
\right)
 \end{eqnarray}
 so our determinant is
\begin{equation}
J_\R=
\varepsilon \prod_{m=1}^M \(\beta_m- \frac{1}{\beta_m}\)
\left| \begin{array}{cccccccc}
e'_{0,1}&e'_{0,2} & \cdots & e'_{0,M}\\
e'_{1,1}&e'_{1,2} & \cdots & e'_{1,M}\\
\vdots&\vdots & \ddots & \vdots
 \\
e'_{M-1,1}&e'_{M-1,2} & \cdots & e'_{M-1,M}\\
\end{array}\right| .
\end{equation}
Here and following we use $\varepsilon$ to denote a number with absolute
value~$1$, which may be different at each occurrence.

 Now we define the following polynomials:
 \begin{equation}
 \begin{split}\label{eqn:flpolys}
 f_{\ell}(x)&=
(x+1)
 \prod_{\ontop{m=1}{m\neq \ell}}^M
 (x-\beta_m)(x-\frac{1}{\beta_m})= \sum_{n=0}^{N-2}(-1)^n e'_{n,\ell} \;x^{N-2-n} \cr
 &= \sum_{n=0}^{M-1}(-1)^n e'_{n,\ell} \;\(x^{N-2-n} + x^{n} \) \cr
 &=
 x^{M-\frac12} \sum_{n=0}^{M-1}(-1)^n e'_{n,\ell} \;\(x^{M-\frac12-n} + x^{-M+\frac12+n}\),
 \end{split}
 \end{equation}
 where the next-to-last step used the fact that
$e'_{n,\ell}= -e'_{N-2-n,\ell}$, since the roots come in
conjugate pairs except for the extra root at~$-1$.

Note that $f_{\ell}(\beta_m) =0$ if $\ell\not = m$. Also we have
\begin{equation}
f_\ell(e^{it}) = 2 e^{i (M-\frac12 )t} \sum_{n=0}^{M-1}(-1)^n e'_{n,\ell}
\cos((M-\tfrac12-n)t) .
\end{equation}
Therefore, with
 \begin{multline}\label{eq:cosinevand}
V(t_1,\ldots , t_M)= \cr
  \begin{vmatrix}
\cos((M-\frac12) t_1)&-\cos((M-\frac32) t_1)  &
     \cdots & (-1)^{M-2}\cos(\frac32 t_1) & (-1)^{M-1}\cos(\frac12 t_1)\\
&&&&\\
\cos((M-\frac12) t_2)&-\cos((M-\frac32) t_2)  &
     \cdots & (-1)^{M-2}\cos(\frac32 t_2) & (-1)^{M-1}\cos(\frac12 t_2)\\
 \vdots & \vdots & \ddots  &\vdots & \vdots \\
\cos((M-\frac12) t_M)&-\cos((M-\frac32) t_M)  &
     \cdots & (-1)^{M-2}\cos(\frac32 t_M) & (-1)^{M-1}\cos(\frac12 t_M)\\
\end{vmatrix}
 \end{multline}
we have
\begin{equation}\label{eqn:VJ}
V(t_1,\ldots, t_M) J_\R = C_M \left| \begin{array}{ccccccc}
f_1(\beta_1)&0  &
  \cdots & 0\\
&&&\\
0&f_2(\beta_2)  &
   \cdots &  0\\
 \vdots & \vdots & \ddots & \vdots  \\
0&0 &
  \cdots &  f_M(\beta_M) \\
\end{array}\right|
 \end{equation}
where
\begin{equation}\label{eqn:CM}
C_M= \varepsilon
2^{-M} \prod_{m=1}^M \(\beta_m- \frac{1}{\beta_m}\) .
\end{equation}

Since
\begin{equation}
2 \sin(\tfrac12 t) \cos((M-\tfrac12-n) t)= \sin((M-n)t - \sin((M-1-n) t),
\end{equation}
by multiplying each row by $\sin(\frac12 t_n)$ we have that
\begin{equation}\label{eqn:sint2factor}
V(t_1,\ldots t_M)
\cdot 2^M
\prod_n
\sin(\tfrac12 t_n)
 \end{equation}
equals a determinant whose rows are
\begin{equation}
\bigl(  \sin(M t_j)  - \sin((M-1) t_j)
\ \ \
\cdots
\ \ \
 \sin(2 t_j) - \sin( t_j )
\ \ \
 \sin(t_j)
\bigr) .
\end{equation}
By elementary column operations starting with the last column,
this
equals a determinant with rows
\begin{equation} \label{eqn:sinevand}
\bigl(  \sin(M t_j)
\ \ \
 \sin((M-1) t_j)
\ \ \
\cdots
\ \ \
 \sin(2 t_j)
\ \ \
 \sin(t_j)
\bigr) .
\end{equation}

Since
\begin{equation}\label{eqn:sinnp1}
\sin((n+1) t)=  \sin(t)\( 2^n\cos^n(t ) + \text{ lower order terms in } \cos(t) \),
\end{equation}
by elementary column operations we find that
the determinant whose rows are~\eqref{eqn:sinevand}
equals a determinant with rows
\begin{equation}\label{eqn:sintfactor}
\sin(t_j)\times
\bigl( 2^{M-1}  \cos^{M-1}(t_j)
\ \ \
 2^{M-2}  \cos^{M-2}(t_j)
\ \ \
\cdots
\ \ \
2 \cos(t_j)
\ \ \
1
\bigr) .
\end{equation}
Since this last determinant is a Vandermonde, we have shown
 \begin{equation}\label{eqn:Vis}
\begin{split}
V(t_1,\ldots t_M)& =
2^{M(M-3)/2}
\prod_n\frac{\sin(t_n)}{\sin(\frac12 t_n)}
\Delta(\cos(t_1),\ldots,\cos(t_M)) \cr
& =
2^{M(M-1)/2}
\prod_n \cos(\tfrac12 t_n)
\Delta(\cos(t_1),\ldots,\cos(t_M)).
\end{split}
\end{equation}

Combining \eqref{eqn:VJ}, \eqref{eqn:CM}, and \eqref{eqn:Vis} we have
 \begin{equation}\label{eqn:jac1a}
J_\R  =\, \varepsilon  2^{-M(M+1)/2}
\frac{\prod_m(\beta_m - \overline{\beta}_m) \prod_{m} f_m(\beta_m)}
        {\prod_m  \cos (\frac12 t_m)\Delta(\cos(t_1),\ldots,\cos(t_M))} .
\end{equation}
Since,
\begin{align}
\prod_{m} f_m(\beta_m) &=
\prod_{m} (\beta_m+1)
\prod_{k\neq m}
(\beta_m-\beta_k)(\beta_m-\overline{\beta}_k) \cr
 & =
2^{2 M^2-M}
\prod_{m} \cos(\tfrac12 t_m) \prod_{k\neq m}
\sin\(\frac{t_m-t_k}{2}\) \sin\(\frac{t_m+t_k}{2}\),
\end{align}
and
\begin{align}
\Delta(\cos t_1,\ldots,\cos t_M)&= \prod_{j<k} (\cos t_k - \cos
t_j) \cr &= 2^{M(M-1)/2} \prod_{j<k}  \sin\(\frac{t_j-t_k}{2}\)
\sin\(\frac{t_j+t_k}{2}\),
\end{align}
and $\beta_m-\overline{\beta}_m=2i\sin(t_m)$,
we find that, 
\begin{align}
J_\R &=
\varepsilon \,2^{M^2}
\prod_m \sin(t_m)
\prod_{j<k}
 \sin\(\frac{t_k-t_j}{2}\) \sin\(\frac{t_k+t_j}{2}\)
\cr
& =
\varepsilon  \prod_m(\beta_m - \overline{\beta}_m)
\prod_{j<k} (\beta_k-\beta_j)(\beta_k-\overline{\beta}_j) ,
\end{align}
as claimed.\qed

In the case of even degree, only a few modifications are needed.
The polynomials \eqref{eqn:flpolys} are replaced by
 \begin{equation}
 \begin{split}
 f_{\ell}(x)&=
 \prod_{\ontop{m=1}{m\neq \ell}}^M
 (x-\beta_m)(x-\frac{1}{\beta_m}) \cr
 &= x^{M-1}\( \sum_{n=0}^{M-2}(-1)^n e'_{n,\ell} \;\(x^{M-1-n} + x^{-M+1+n}\)
+ (-1)^{M-1} e'_{M-1,\ell} \), \end{split}
 \end{equation}
so
\begin{equation}
 f_{\ell}(e^{i t})=
\varepsilon
\(
\,2 \sum_{n=0}^{M-2}(-1)^n e'_{n,\ell} \cos((M-1-n)t)
+ (-1)^{M-1} e'_{M-1,\ell}
\)
.
\end{equation}
The determinant $V(t_1,\ldots,t_M)$
has entries $\cos((M-1-n)t_j)$, except for the last column,
where
$n=M-1$, which is
multiplied by a factor of~$\frac12$.
Since
\begin{equation}
\cos(n t)=   2^{n-1}\cos^n(t ) + \text{ lower order terms in } \cos(t) ,
\end{equation}
we recognize $V(t_1,\ldots,t_M)$ as a Vandermonde,
which is $2^{M}$ times smaller than the Vandermonde which appeared in the
odd degree case.
The only other differences
in the calculation is to
omit the
factor $2^M \prod \sin(t_n/2)$ from \eqref{eqn:sint2factor} and
$\sin(t_j)$ from \eqref{eqn:sintfactor}.   The overall effect
of those factors is to multiply the Jacobian by
\begin{equation}
\prod_n \frac{\sin(t_n)}{2\sin(\frac{t_n}{2})}
=
\prod_n \cos\left(\frac{t_n}{2}\right).
\end{equation}
That factor replaces $\prod_m(\beta_m+1)=2^M\prod_m \cos(\frac12 t_m)$ which is now omitted
from~$\prod_m f_m(\beta_m)$, the power of $2$ making up for the factor
of $2^M$ missing from~$V(t_1,\ldots,t_M)$.  So the end result is the exact same formula
for the Jacobian.

\renewcommand{\theequation}{A\arabic{equation}}
\setcounter{equation}{0}
\section*{Appendix A. Epstein zeta functions associated to binary quadratic forms.}

An example ensemble of Dirichlet series with functional equation
is the Epstein zeta-functions associated to a binary quadratic
form. These functions have functional equations but no Euler
product. Indeed it is conjectured (Sarnak's rigidity conjecture)
that there does not exist a nontrivial continuous family of
Dirichlet series with functional equation and Euler product.

To create an Epstein zeta-function, begin with a binary quadratic
form
\begin{equation}
Q(m,n)=a m^2+ b m n + c n^2.
\end{equation}
We assume that $Q$ is positive-definite, meaning that $Q(m,n)>0$
if $(m,n)\neq (0,0)$.  This is equivalent to $a>0$ and
$\Delta=b^2-4 a c<0$. In number theory one considers the case of
$a,b,c\in \mathbb Z$, but we will allow $a,b,c$ to be arbitrary
real numbers.

The Epstein zeta-function associated to $Q$ is defined as
\begin{equation}\label{eqn:LQ}
L_Q(s) := {\sum_{m,n}}' \frac{1}{Q(m,n)^s},
\end{equation}
where the notation $\sum'$ means that we sum over all pairs of
intgers $(m,n)$ except for $(0,0)$.  This series converges for
$\Re(s)>1$, and $L_Q(s)$ has a meromorphic continuation to the
entire complex plane except for a simple pole at~$s=1$. The
Epstein zeta-function satisfies the functional equation
\begin{equation}
\left(\frac{\sqrt{|\Delta|}}{2\pi}\right)^s \Gamma(s) L_Q(s) =
\left(\frac{\sqrt{|\Delta|}}{2\pi}\right)^{1-s} \Gamma(1-s)
L_Q(1-s),
\end{equation}
where $\Gamma(s)$ is the Gamma function. Thus, $L_Q(s)$ has a
functional equation similar to that of the Riemann
$\zeta$-function and other $L$-functions of number theory. For
certain quadratic forms the Epstein zeta-function can be expressed
in terms of $L$-functions.  For example,
\begin{equation} \label{eqn:m2n2}
L_{m^2+n^2}(s) = 4 \zeta(s)L(s,\chi_4),
\end{equation}
where $\zeta(s)$ is the Riemann zeta-function and $L(s,\chi_4)$ is
the Dirichlet $L$-function for the nontrivial character
modulo~$4$. This Epstein zeta function is expected to satisfy the
Riemann hypothesis, but note that its zeros are a superposition of
two independent sets of zeros and so will not display the
characteristic quadratic repulsion expected for $L$-functions with
functional equation and Euler product. For the Epstein zeta
function~\eqref{eqn:m2n2}, and a handful of other special cases,
$L_{Q}(s)$ has an Euler product.  In the case that $a,b,c$ are
integers, $L_{Q}(s)$ is a finite linear combination of Dirichlet
$L$-functions. However, for generic $Q(m,n)$ there is no
arithmetic structure to the Epstein zeta-function $L_Q(s)$, and
hence no Euler product, and it is these ``random Dirichlet series
with functional equation'' which illustrate that we are not
studying an empty set of functions.


There is a natural way to parametrize the Epstein zeta-functions.
Write
\begin{equation}
L_Q(s)=a^{-s} {\sum_{m,n}}' \frac{1}{|m+z n|^{2s}},
\end{equation}
where $z=x+i y$ is a complex number satisfying $|z|^2=c/a$ and $2
x=b/a$. Note that the conditions on $a,b,c$ guarantee that such a
number $z$ exists, and $z$ is uniquely determined if we also
assume~$y>0$. The above formula for $L_Q$ allows us to relate the
Epstein zeta-function to an Eisenstein series on $SL(2,\mathbb
Z)$:
\begin{equation}
E(z;s):= \frac12 {\sum_{m,n}}' \frac{y^s}{|m z+ n|^{2s}}.
\end{equation}
Specifically, we have $L_Q(s)=2 (a y)^{-s} E(z;s)$.

The Eisenstein series $E(z;s)$ is a well-understood object from
number theory. See \cite{kn:Iwaniec02} for details.  However, in
that theory one usually fixes $s$ and considers $E(z;s)$ as a
function of~$z$. Indeed, as a function of $z$ it has many
fascinating properties: it is an eigenfunction of the hyperbolic
Laplacian, and it is invariant (up to an automorphic factor) under
the linear-fractional action of $SL(2,\mathbb Z)$ on the upper
half-plane. Thus, the Epstein zeta-functions associated to a
binary quadratic form are parametrized by the points in a
fundamental domain for $SL(2,\mathbb Z)$:
\begin{equation}
\mathcal F = \left\{z=x+i y\ :\ -\frac12<x\le \frac 12 ,\ y>0,\
|z|>1 \right\}.
\end{equation}

In \cite{kn:fk} evidence is given that such functions can be
modeled by random trigonometric polynomials.


\newpage
\end{document}